\documentclass[10pt,conference,a4paper]{IEEEtran}

\usepackage{times}
\usepackage{epsfig}
\usepackage{graphicx}
\usepackage{amsmath}
\usepackage{amssymb}
\usepackage{subfig}
\usepackage{lipsum}

\usepackage[pagebackref=true,breaklinks=true,letterpaper=true,colorlinks,bookmarks=false]{hyperref}

\begin{document}

\title{ISP4ML: The Role of Image Signal Processing in Efficient Deep Learning Vision Systems}



%
\author{\IEEEauthorblockN{Patrick Hansen\IEEEauthorrefmark{1},
Alexey Vilkin\IEEEauthorrefmark{2},
Yury Khrustalev\IEEEauthorrefmark{1},
James Imber\IEEEauthorrefmark{3},
Dumidu Talagala\IEEEauthorrefmark{1}
\\
David Hanwell\IEEEauthorrefmark{1},
Matthew Mattina\IEEEauthorrefmark{1} and
Paul N. Whatmough\IEEEauthorrefmark{1}}
\IEEEauthorblockA{\IEEEauthorrefmark{1}Arm Research, \{first.last\}@arm.com}
\IEEEauthorblockA{\IEEEauthorrefmark{2}Twitter, alexey.vilkin1@gmail.com}
\IEEEauthorblockA{\IEEEauthorrefmark{3}Imagination Technologies, james.imber@imgtec.com}}


\maketitle

\begin{abstract}
Convolutional neural networks (CNNs) are now widely deployed in a variety of computer vision (CV) systems.
These systems typically include an image signal processor (ISP), even though the ISP is traditionally designed to produce images that look appealing to humans.
In CV systems, it is not clear what the role of the ISP is, or if it is even required at all for accurate prediction.
In this work, we investigate the efficacy of the ISP in CNN classification tasks and outline the system-level trade-offs between prediction accuracy and computational cost.
To do so, we build software models of a configurable ISP and an imaging sensor to train CNNs on ImageNet with a range of different ISP settings and functionality.
Results on ImageNet show that an ISP improves accuracy by 4.6\%-12.2\% on MobileNets. Results from ResNets demonstrate these trends also generalize to deeper networks.
An ablation study of the various processing stages in a typical ISP reveals that the tone mapper is the most significant stage when operating on high dynamic range (HDR) images, by providing 5.8\% average accuracy improvement alone.
We also show that the ISP increases the generalization of CNNs across two different image sensors by a significant 17.5\%.
Overall, the ISP benefits system efficiency because the memory and computational costs of the ISP is minimal compared to the cost of using a larger CNN to achieve the same accuracy.
\end{abstract}

\section{Introduction}
\label{sec:intro}

Deep convolutional neural networks (CNNs) have surpassed traditional algorithms in terms of accuracy on many computer vision (CV) tasks.
In many emerging CV-oriented applications, such as advanced driver assistance systems (ADAS), camera sensor data in the system is consumed solely by the CV processing algorithms and is never viewed by humans at all\footnote{Perhaps with the exception of debugging and development purposes.}. 
Therefore, the image representation is no longer constrained to enable human viewing.
This is in stark contrast to applications such as photography, where the goal is to generate images that are subjectively aesthetically pleasing to the human eye, as well as requiring an 8-bit, gamma-corrected, RGB format to be properly output on a typical display. 
Without the need to conform to these traditional constraints, there is scope to improve the computational efficiency and algorithmic performance of the system by optimizing the image representation. 

\begin{figure}[t]
\centering
\vspace{-10pt}
\subfloat[Original RGB]{\centering\label{fig:tiger-vanilla}\includegraphics[width=0.14\textwidth]{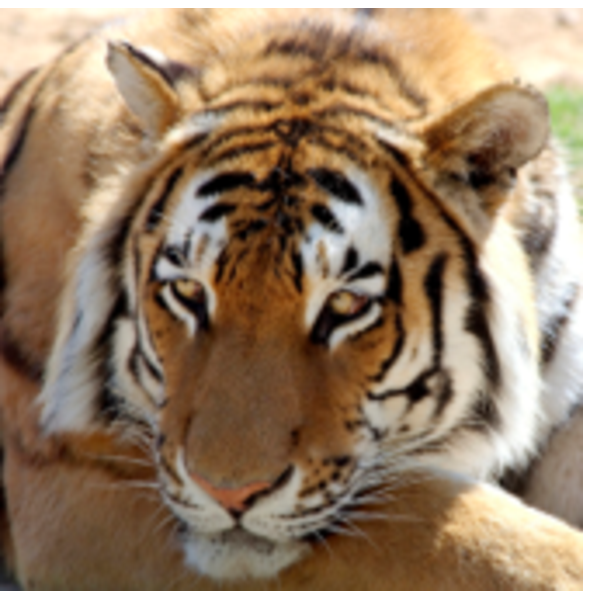}}
\hspace{0.01\textwidth}
\subfloat[Simulated Raw]{\centering\label{fig:tiger-raw}\includegraphics[width=0.14\textwidth]{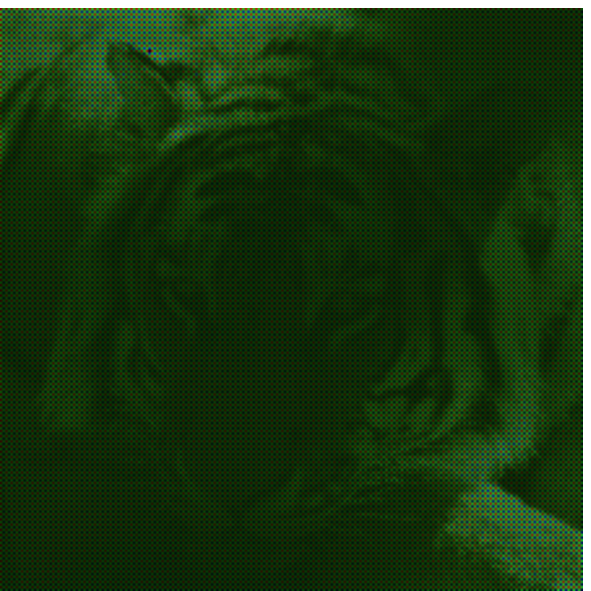}}
\hspace{0.01\textwidth}
\subfloat[Simulated RGB]{\centering\label{fig:tiger-rgb}\includegraphics[width=0.14\textwidth]{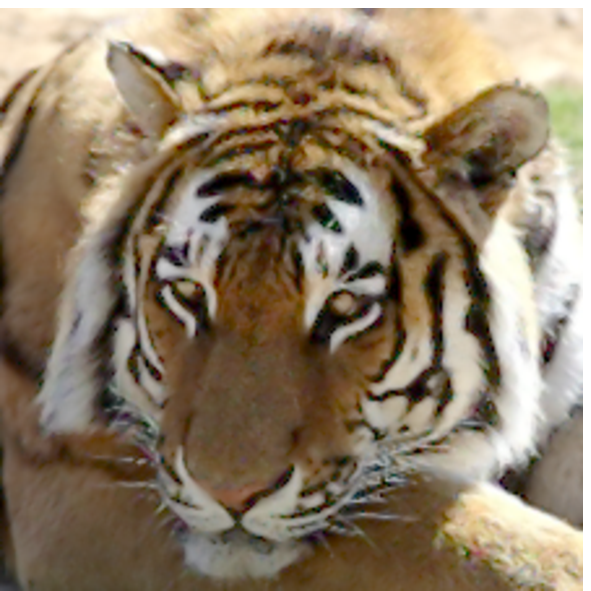}}
\vspace{-5pt}
\caption{Example ImageNet sample: (a) original (RGB), (b) processed by capture model to simulate raw HDR sensor data (colorized), and (c) output of ISP software model operating on simulated raw data.}
\end{figure}

\begin{figure*}[t]
\centering
\vspace{-15pt}
\includegraphics[width=0.75\textwidth]{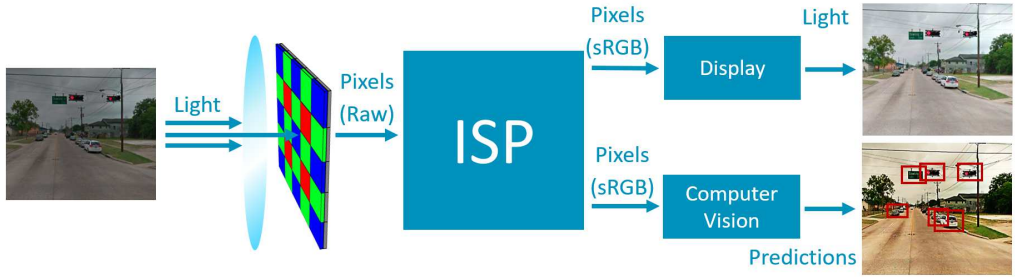}
\vspace{-15pt}
\caption{Typical computer vision pipeline.}
\label{fig:landscape}
\end{figure*}

CNNs are typically trained on datasets such as ImageNet~\cite{imagenet}, which contain images processed using an image signal processor (ISP) and stored in RGB image representation. An ISP is a hardware component consisting of several pipelined processing stages designed to process raw Bayer color filter array (CFA) sensor data into RGB output images.
Figure \ref{fig:landscape} depicts how an ISP is typically used in embedded CV. A camera lens focuses light onto a CFA image sensor, which produces a single plane of digital raw pixel values (Figure \ref{fig:tiger-raw}). 
These raw images are then processed by an ISP to generate RGB images (Figure \ref{fig:tiger-rgb}), which are consumed either by a display for human viewing or by a CV algorithm.



While optimization of the traditional image display path in Figure~\ref{fig:landscape} has been studied, the role and optimization of the ISP for CNN inference is relatively unexplored.
Traditionally, ISP algorithms are designed to optimize for metrics such as peak signal-to-noise-ratio (PSNR), modulation transfer function (MTF), and subjective human rating scores.
However, these metrics clearly may not be optimal for CNN prediction accuracy.
If we could reduce the complexity of the ISP by removing unneeded functions, that would lead to lower latency and lower energy consumption.
Even though CNNs are exceptional in their ability to adapt to various input representations, co-optimization of the ISP and network architecture may also enable reduction in the size of the CNN model at the same prediction accuracy, again reducing latency and energy.
Any complexity savings that can be made to the CNN are especially potent, as CNNs have high computational cost, even when implemented on highly-efficient and high-throughput neural processing units (NPUs)~\cite{smiv,sta2020cal}. 

In this paper, we seek to understand the impact of the ISP on CNN classification problems. 
However, this is not straightforward, because standard labeled datasets commonly used for CNN experiments consist of RGB images, which have already been processed from raw sensor data with fixed ISP settings.
Collecting and labeling a new dataset of raw images for training is prohibitively expensive for a useful number of images.
Instead, we follow the approach of Brooks et al.~\cite{brooks} to simulate RAW images from RGB using a model of an imaging sensor.
We generate RGB images from simulated raw images using a configurable, industry-grade, high dynamic range (HDR) ISP.
This enables us to generate datasets with a variety of ISP configurations and settings. We evaluate state-of-the-art CNN model architectures trained on these datasets to formulate hypotheses on the impact of ISP design on CNN classification accuracy.

Our results confirm that there is indeed a benefit, especially marked for compact models, to incorporating an ISP in CNN inference. 
Furthermore, we determine what functions of the ISP are most significant to achieve accurate predictions. 
Finally, we present a system-level analysis of the ISP cost trade-offs to demonstrate the efficiency benefits of an ISP. 

In summary, this paper makes the following contributions:
\begin{itemize}
    \item \textbf{Raw vs RGB.} 
    Using a capture model and an industry-grade ISP model, we process ImageNet to train and test CNN classification models with configurable ISP settings.
    Results on MobileNets show an average 7.0\% accuracy improvement from full ISP processing over raw images.
    \item \textbf{Impact of Model Size.}
    The accuracy improvement provided by ISP processing is most significant for compact models, suggesting larger models are more readily able to learn from raw images. 
    Experiments with larger models, ResNet-50 and ResNet-101, further support this claim.
    \item \textbf{ISP Stage Ablation.}
    The tone mapper has the single greatest impact on accuracy for models trained on HDR images, providing 5.8\% average improvement alone. Results indicate this is because CNNs struggle to learn from heavily skewed pixel distributions.
    \item \textbf{Compute and Memory Analysis.}
    Using an ISP actually \textit{improves} overall system hardware efficiency compared to a baseline with no image processing.
    This result can be attributed to the negligible cost of the ISP in comparison to relatively expensive CNN inference.
    It also improves generalization to different imaging sensors in the hardware platform.
\end{itemize}
    

The remainder of the paper is organized as follows: Section \ref{sec:related} describes related work, Section \ref{sec:method} gives background and methodology. Results are presented in Section \ref{sec:stages}, and Section \ref{sec:tone-mapping} investigates the impact of tone mapping. Section \ref{sec:system} evaluates trade-offs in accuracy and compute.  Finally, Section \ref{sec:conclusion} concludes the paper.


\section{Related Work}
\label{sec:related}

\textbf{ISP Optimization for CV.}
Recent research and commercial products have begun exploring ISP specialization for vision systems.
For example, Arm Mali-C71~\cite{mali-c71} is an ISP designed specifically for use in ADAS systems.
Liu et al.~\cite{liu} proposed an ISP that can selectively disable stages based on application needs. Lubana et al.~\cite{lubana} proposed a minimal in-sensor accelerator to enable inference directly on sensor outputs without CNN retraining.
Brooks et al.~\cite{brooks} introduced a methodology involving using an imaging sensor model to simulate raw images from RGB images.
Buckler et al.~\cite{buckler} investigated the impact of ISP stages on different CV algorithms, and proposed a sensor design with adjustable precision and subsampling.

Our work follows a similar methodology to Buckler et al.~\cite{buckler} and Brooks et al.~\cite{brooks}, using a sensor model and a configurable ISP model to enable training models with a different input representations. 
We extend the prior work by: 1) modelling a product-quality high dynamic range (HDR) ISP, 2) validating our imaging sensor model using a dataset of raw images that we captured for our experiments, and 3) analyzing the impact the ISP on compute efficiency.
Additionally, we correct and improve upon some combinations of individual ISP stages used in Buckler et al. ~\cite{buckler}, e.g., color correction on Bayer images.
We are also the first to investigate the impact of the ISP on training CNNs on ImageNet scale problems, using both highly-efficient compact models (MobileNet family) and highly-accurate deep models (ResNet family).


\textbf{Neural ISP Algorithms.}
CNNs have been developed for denoising~\cite{denoise-cnn}, demosaicing~\cite{demosaicnet}, and end-to-end imaging solutions~\cite{deepisp}. 
These works demonstrate that CNNs are capable of replicating ISP functions, and even improving on them in terms of PSNR metrics.
However, this improved performance comes at a significant compute cost: on the order of 1M operations/pixel~\cite{deepisp}, compared to around 1K ops/pixel for a traditional ISP.
Diamond et al.~\cite{dirty-pixels} investigates the impact of image distortions on classification accuracy, and propose a network architecture for joint deblurring, denoising, and classification. 
Our work differs from previous works in that we do not attempt to emulate ISP functions with neural algorithms, but rather explore the need for ISP functions in the first place.

\textbf{Efficient CNN Inference.}
CNNs are computationally expensive, so it is essential to have both efficient hardware and an efficient network architecture.
Recent research has focused on developing dedicated hardware for efficient deep learning inference~\cite{nvdla, eyeriss}.
Additionally, significant progress has been made developing efficient CNN architectures by hand~\cite{mobilenetv1, efficientnet, squeezenet}, or using search techniques~\cite{darts}. In this work, we focus on MobileNets~\cite{mobilenetv1}, a family of CNNs designed for energy-efficient inference. 
State-of-the-art models in both the highly-efficient~\cite{mobilenetv3} and highly-accurate~\cite{efficientnet} domains build upon the basic MobileNet structure. 
We also use ResNets~\cite{resnet} to demonstrate generalization of our results to larger networks.

\textbf{Hardware Feature Extractors for CNNs.}
When used in an CV system, an ISP can be considered a hardware feature extractor. Whatmough et al.~\cite{fixynn} demonstrated efficiency benefits from fully-fixed hardware CNN feature extraction. Other works have demonstrated energy improvement from in-sensor analog feature extraction~\cite{analog}. Our work demonstrates similar benefits for handcrafted ISP features.

\section{Background and Methodology}
\label{sec:method}

\begin{figure}[t]
\centering
\vspace{-10pt}
\subfloat[ISP Model]{\centering\label{fig:isp}\includegraphics[width=0.15\textwidth]{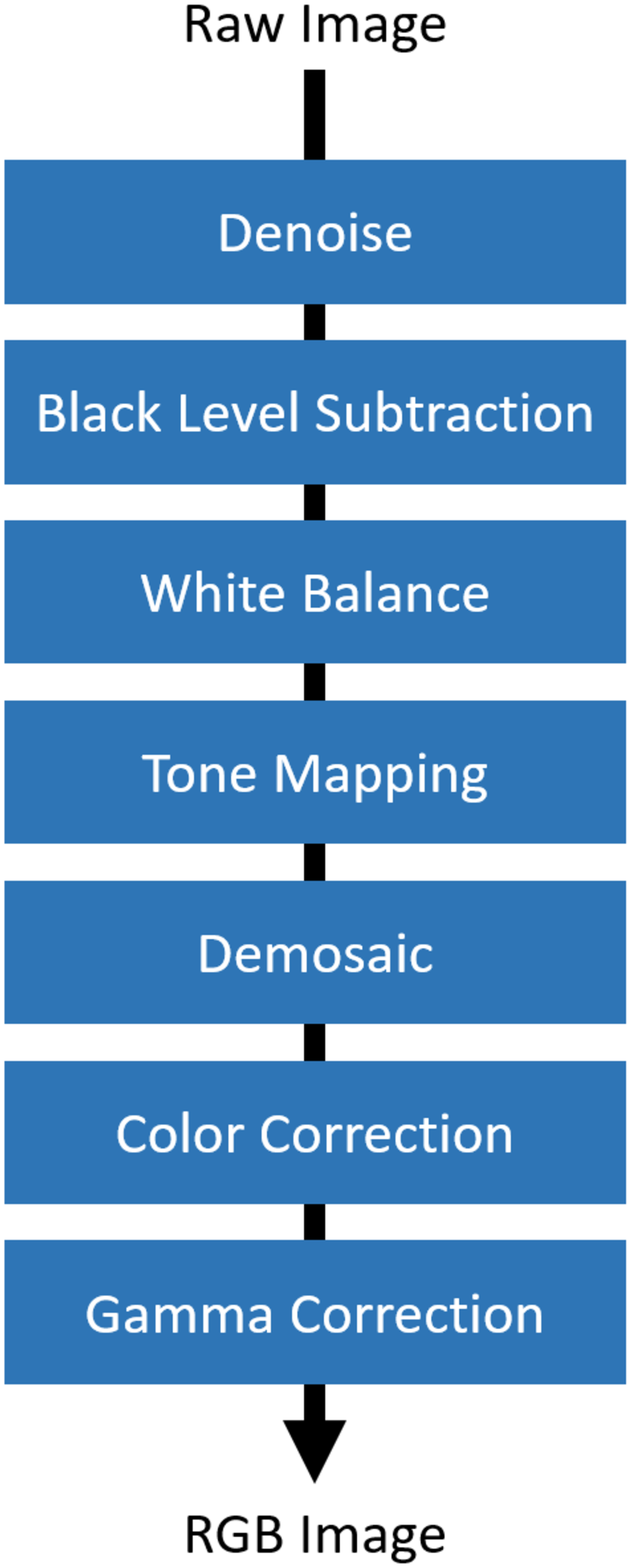}}
\hspace{20pt}
\subfloat[Capture Model]{\centering\label{fig:capture}\includegraphics[width=0.15\textwidth]{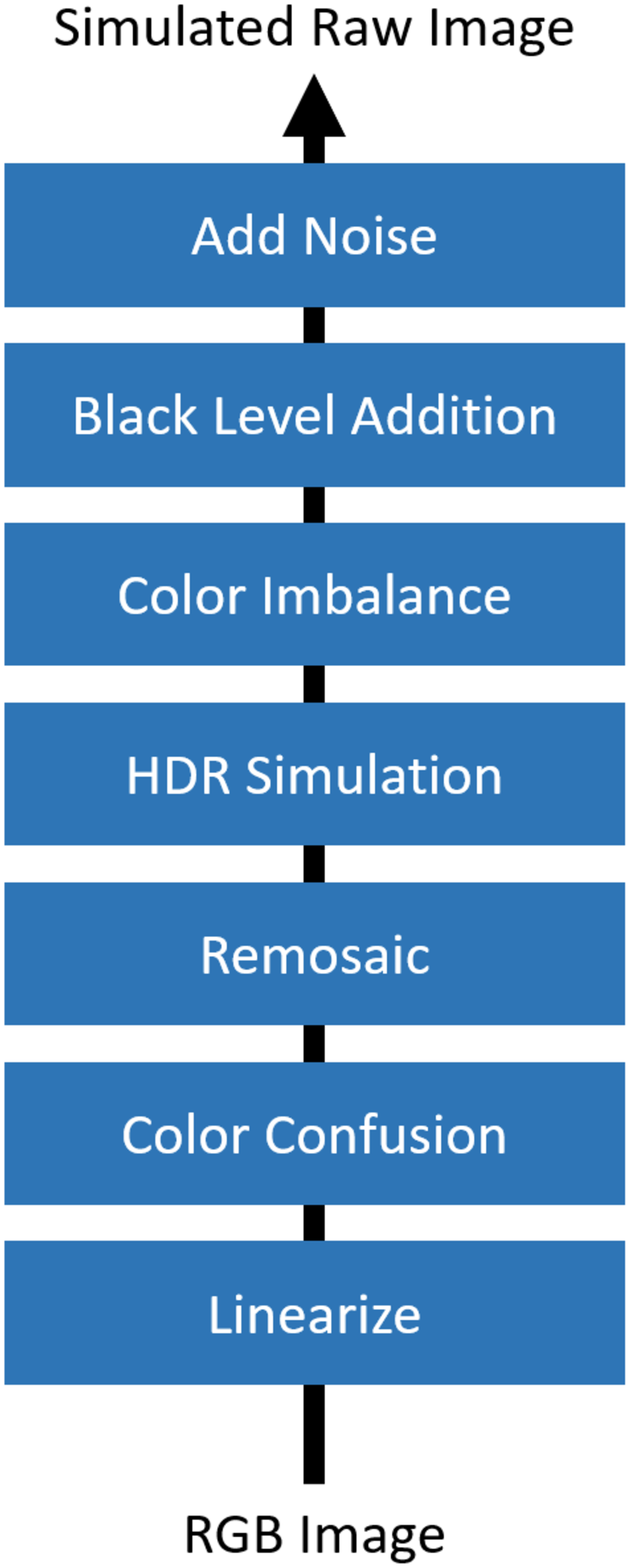}}
\caption{Key components of the ISP and capture models.}
\end{figure}

To evaluate the impact of the ISP on CV systems, we trained CNNs on images processed by a variety of ISP configurations (including no ISP at all). 
This approach enables us to study the impact of each ISP stage on prediction accuracy on a large dataset (e.g. ImageNet). 
In this section, we describe the ISP we model, our approach in generating images for training/testing, and the benchmarks used.

\subsection{ISP Model Overview}
\label{sec:method:isp}
For our experiments, we implement the ISP stages shown in Figure \ref{fig:isp} in a C++ software model using production-grade algorithms.
Although some implementation details are proprietary, we provide brief descriptions of each stage below:
\begin{itemize}
    \item \textbf{Denoise}: based on a separable bilateral filter.
    \item \textbf{Black level subtraction}: simple subtraction of a constant.
    \item \textbf{White balance}: global scaling of each color channel by gain values (derived from sensor measurements).
    \item \textbf{Tone mapping}: based on local histogram equalization.
    \item \textbf{Demosaic}: a directional demosaic similar to \cite{demosaic}.
    \item \textbf{Color correction}: pixel-wise multiplication by a color-correction matrix (derived from sensor measurements).
    \item \textbf{Gamma}: apply a standard sRGB gamma curve \cite{srgb}.
\end{itemize}
The ISP software model allows stages to be optionally enabled or disabled\footnote{Limited to valid combinations with compatible data representations.}. The ISP design itself is not the main focus of this work; rather, it allows us to test a variety of ISP configurations in the ISP design space in relation to CNN accuracy.

\subsection{Capture Model for Simulating Raw Data}
\label{sec:method:raw}
To evaluate the impact of ISP design decisions on CNN model performance, we require a large quantity of labelled data. 
However, datasets relevant to CNN tasks are nearly exclusively captured in 8-bit RGB format. 
In other words, these images have already been processed by an ISP, without storing the original raw image data.
Even worse, these images are typically captured using a range of different (unknown) sensors and ISPs.
Although some raw image datasets do exist, they are designed for research on ISP demosaicing and denoising algorithms~\cite{raise,mcmaster} and are not suitable for training modern CNNs, as they are too small and unlabeled. 

As such, we simulate raw HDR captures by processing standard RGB image datasets using a software model which we call the \textit{capture model}. This can be thought of as a ``reverse ISP'' in the sense that it generates raw images from RGB images.
However, the goal of the capture model is to generate images with statistics representative of raw images, rather than to perfectly recreate the original raw image.
The capture model accomplishes this by applying an approximation of the inverse\footnote{Not all ISP stages are perfectly invertible, \textit{viz.} denoise, white balance, and tone mapping. The original sensor noise cannot be perfectly recreated from a denoised image. Similarly, white balance and tone mapping use original image statistics that cannot be recovered after-the-fact.} of each stage in the ISP pipeline. Figure \ref{fig:capture} depicts the approach.

The capture model simulates a HDR scene by applying an inverse tone mapping function to a standard dynamic range RGB image (Figure \ref{fig:tiger-vanilla}) and captures 3 snapshots of that HDR scene at different levels of exposure (similar to a sensor in practice). The capture model thereafter processes these images separately introducing colour confusion, imbalance, noise, etc. to generate raw HDR captures. The 3 raw images are subsequently combined into a single RAW HDR image (Figure \ref{fig:tiger-raw}) by exposure alignment prior to ISP processing.

Below, we briefly summarize each stage of the capture model:
\begin{itemize}
    \item \textbf{Linearize}: apply the inverse of sRGB gamma.
    \item \textbf{HDR simulation}: apply a smooth parameterized curve to spatially modulate image intensity with the aim of recreating a distribution representative of real HDR captures. 
    \item \textbf{Color confusion}: pixel-wise multiplication by a color-confusion matrix (inverse of the color-correction matrix).
    \item \textbf{White imbalance}: global scaling of each color channel by gain values (inverse of color correction gains).
    \item \textbf{Remosaic}: apply a Bayer pattern mask and collapse to a single plane, discarding two-thirds of color data.
    \item \textbf{Black level addition}: simple addition of a constant.
    \item \textbf{Noise addition}: add both Poisson (to simulate shot noise) and Gaussian (to simulate dark current and read noise) noise. 
\end{itemize}
Our capture model bares similarity to the raw image pipeline proposed by Brooks et al.~\cite{brooks}, however our model differs in that our HDR simulation is spatially varying, and we linearize before HDR simulation.
Parameters for color confusion, white imbalance, black level and noise are calibrated on actual sensor measurements. We use the Sony IMX290~\cite{imx290} sensor unless stated otherwise.
We validate our results by testing our trained CNN models on true raw images and observing consistent trends in prediction accuracy.

\subsection{Benchmark CNNs and Datasets}
\label{sec:method:benchmarks}
We trained MobileNet~\cite{mobilenetv1} and ResNet~\cite{resnet} CNN architectures on ImageNet~\cite{imagenet} to evaluate the impact of the ISP. MobileNet is a family of parameter-efficient, scalable CNN architectures, with hyperparameters to scale input resolution and the width of each layer\footnote{We used 224x224 images and width multipliers [0.25, 0.50, 0.75, 1.00].}. 
MobileNets are designed for compute-constrained devices, however derivative network architectures have also demonstrated SOTA results in other application domains~\cite{efficientnet}. 
We use this model architecture for our investigations because MobileNet-like models are highly relevant in the field today, and because its scalability enables us to test on a range of model sizes. To demonstrate the generalization of our results to deeper models, we also train ResNets~\cite{resnet} on ImageNet.


\begin{figure}[t]
\centering
\includegraphics[width=0.4\textwidth]{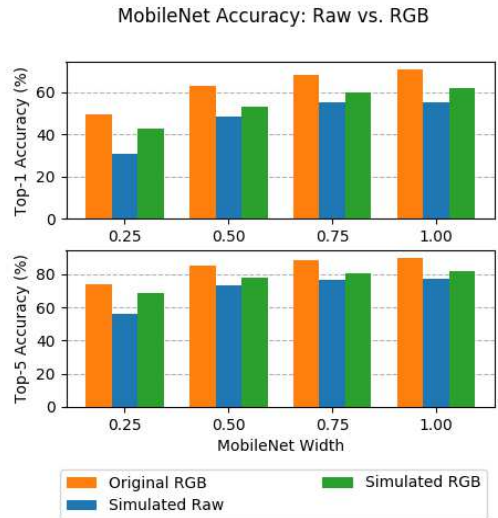}
\vspace{-8pt}
\caption{Raw vs RGB test accuracy for MobileNet on ImageNet. 
ImageNet images (\textit{Original RGB}) are processed using the capture model to generate raw images (\textit{Simulated Raw}), which are then processed using our software ISP model to generate RGB images (\textit{Simulated RGB}).}
\label{fig:raw-vs-rgb}
\vspace{1.5ex}
\end{figure}


\begin{figure}[t]
\centering
\vspace{-8mm}
\includegraphics[width=0.4\textwidth]{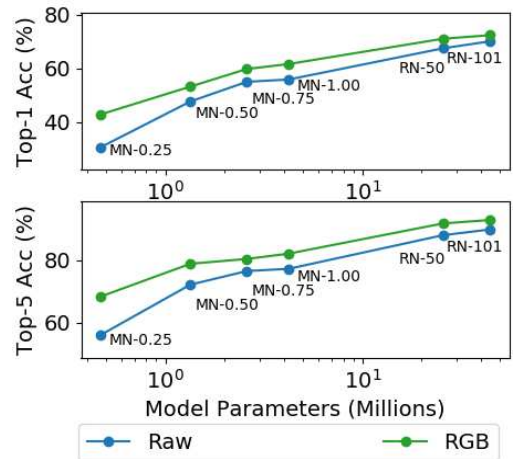}
\vspace{-5mm}
\caption{Raw vs RGB accuracy difference for a range of models (0.47M-44.5M parameters). MN--MobileNet, RN--ResNet.}
\label{fig:delta}
\end{figure}

\begin{figure*}[t]
\centering
\subfloat[MobileNet accuracy for raw and RGB with/without denoise. ]{\centering\label{fig:denoise}\includegraphics[width=0.4\textwidth]{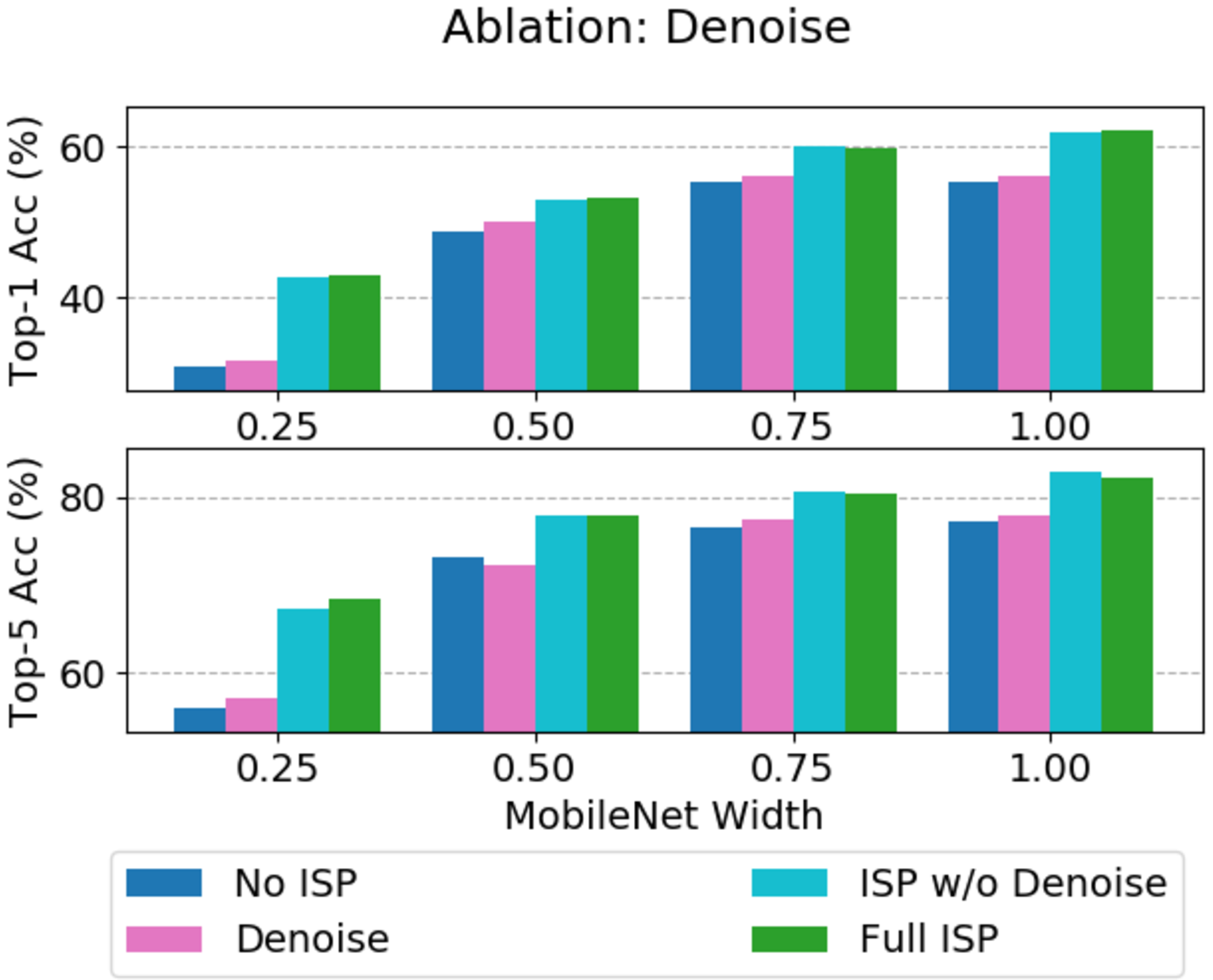}}
\subfloat[MobileNet accuracy over a variety of ISP stages.]{\centering\label{fig:partial}\includegraphics[width=0.4\textwidth]{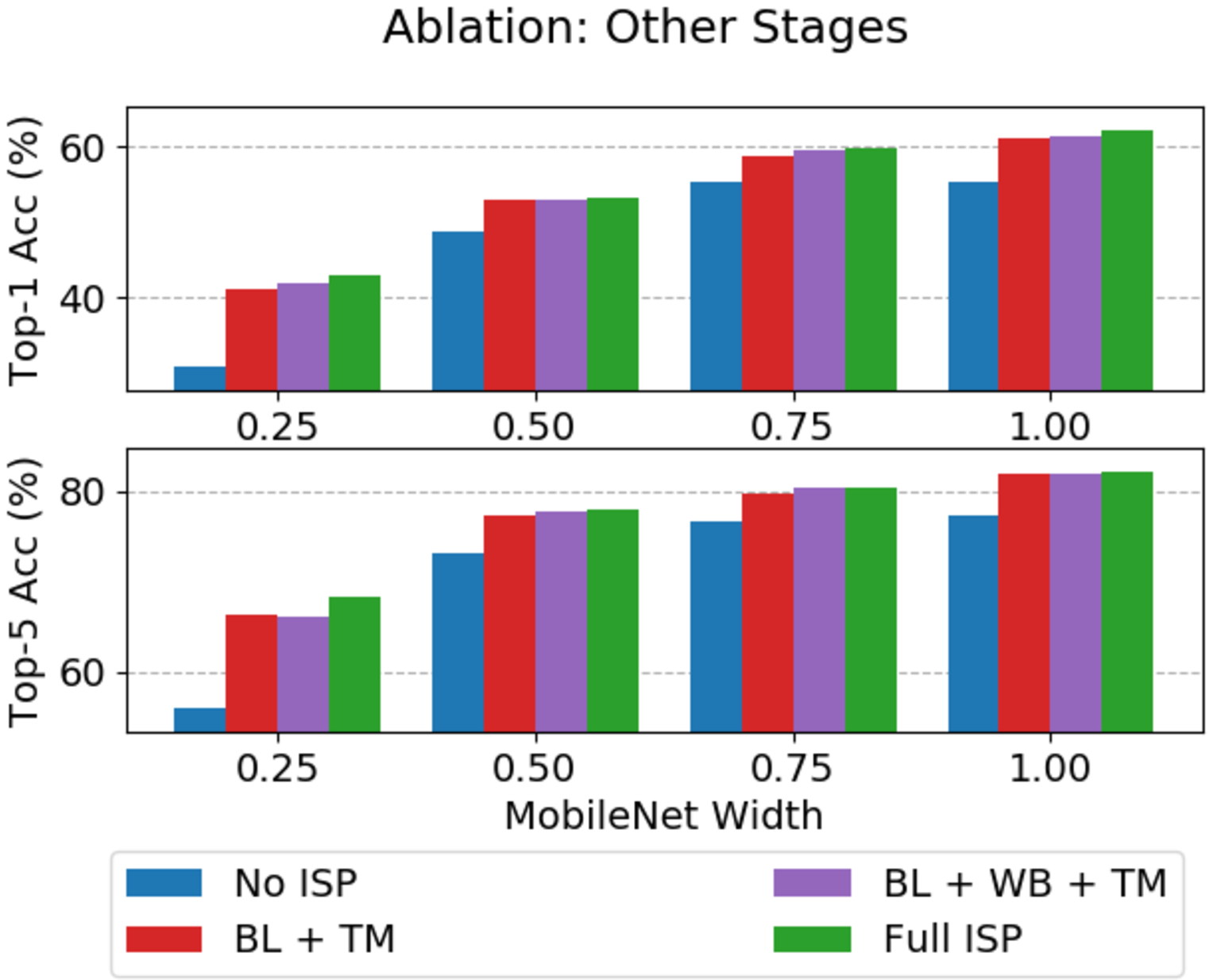}}
\caption{Ablation study of ISP stages. Raw images (baseline) are generated from ImageNet using the capture model directly. The other image sets are generated by processing raw images with a differing number of ISP stages.}
\label{fig:ablation}
\end{figure*}

In our experiments, we train each network with the same training parameters as used in their original publication~\cite{mobilenetv1, resnet}. The only exceptions to this are: (1) batch normalization decay, which we set to 0.99 for models trained on certain image sets, and (2) data augmentation, which we limit to mean and variance normalization for fair comparison between different image representations\footnote{Random cropping, flipping, and certain color augmentations cannot be applied to raw images because they do not preserve the Bayer mosaic pattern.}.
The original training recipes for these networks were carefully tuned for RGB images, so we made sure to tune our training recipe for raw images.

In addition to the training recipes being optimized for RGB images, we acknowledge that the model architectures themselves are tuned for RGB images. 
It may be possible to design network architectures better suited for raw images, however this is a non-trivial question we leave as future work.

\section{Impact of ISP on CNN Accuracy}
\label{sec:stages}


In this section, we discuss the results of CNN training experiments using images generated by different ISP configurations. We consider ISP configurations with all stages enabled, no stages enabled, and stages selectively enabled/disabled.

\subsection{Raw vs. RGB Images}
We first evaluate the overall impact of the ISP on CNN classification accuracy. 
To accomplish this, we compare the performance of models trained on raw images (no ISP processing) against those trained on RGB images (full ISP processing). 
We train MobileNets on two processed versions of ImageNet: one containing simulated raw images generated by the capture model (Section~\ref{sec:method:raw}), and the other containing RGB images generated by the ISP model (Section~\ref{sec:method:isp}) processing the simulated raw images. Figure \ref{fig:raw-vs-rgb} provides the results of this experiment, and compares the accuracy of these models to the published MobileNet results~\cite{mobilenetv1}. This figure demonstrates that the models trained on RGB images outperform those trained on raw images by an average of 7.0\%.

Note that the accuracies of our simulated RGB models are lower than published results (original RGB). This is expected because the capture model cannot perfectly reproduce the original raw images that produced the ImageNet images -- the original color and noise profiles of the original raw images are unknown, and the capture model throws away data by remosaicing and adds noise. Additionally, as noted previously (Section~\ref{sec:method:benchmarks}), we forego data augmentation techniques such as random cropping and flipping for fair comparison between raw and RGB training experiments.

\begin{table}[t]
\centering
\resizebox{\columnwidth}{!}{
\begin{tabular}{c c c c c}
\hline
Model & Params & Image type & Top-1 acc. & Top-5 acc. \\
\hline\hline
ResNet-50 & 25.6M & Raw & 67.45 & 88.04 \\
 & & RGB & 71.01 & 91.82 \\ 
ResNet-101 & 44.5M & Raw & 70.07 & 89.90 \\
 & & RGB & 72.34 & 92.93 \\
\hline
\end{tabular}
}
\caption{Raw vs. RGB accuracies for ResNet-50 and 101.
}
\label{table:resnet}
\end{table}


We notice that the accuracy difference between no ISP processing and full ISP processing is larger for compact models.
Since even the largest MobileNet is a relatively small model, we also repeated this experiment on two sizes of ResNets~\cite{resnet}, to understand the impact of the ISP on larger and deeper models. 
Similar to MobileNet, ResNet experiments (Table~\ref{table:resnet}) show a benefit to training models on RGB images, however the gap in accuracy is smaller seen for MobileNet (c.f. Figure~\ref{fig:delta}). This suggests that larger (or deeper) models, such as ResNets, are more readily able to learn from image representations other than RGB.

\begin{figure*}[t]
\centering
\subfloat[HDR linear raw]{\centering\label{fig:raw-hist}\includegraphics[width=0.32\textwidth]{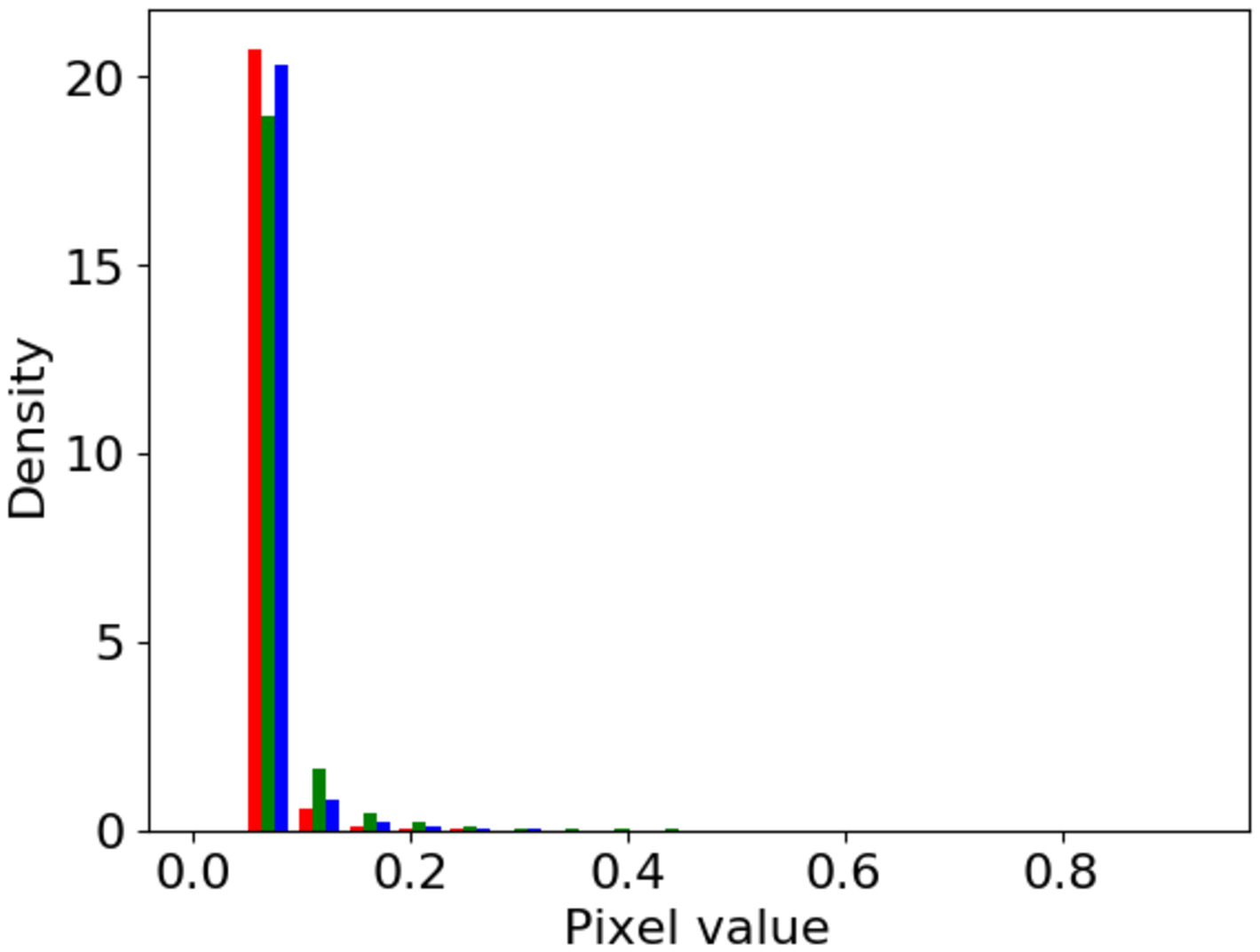}}
\subfloat[Raw after tone-mapping]{\centering\label{fig:iridix-hist}\includegraphics[width=0.32\textwidth]{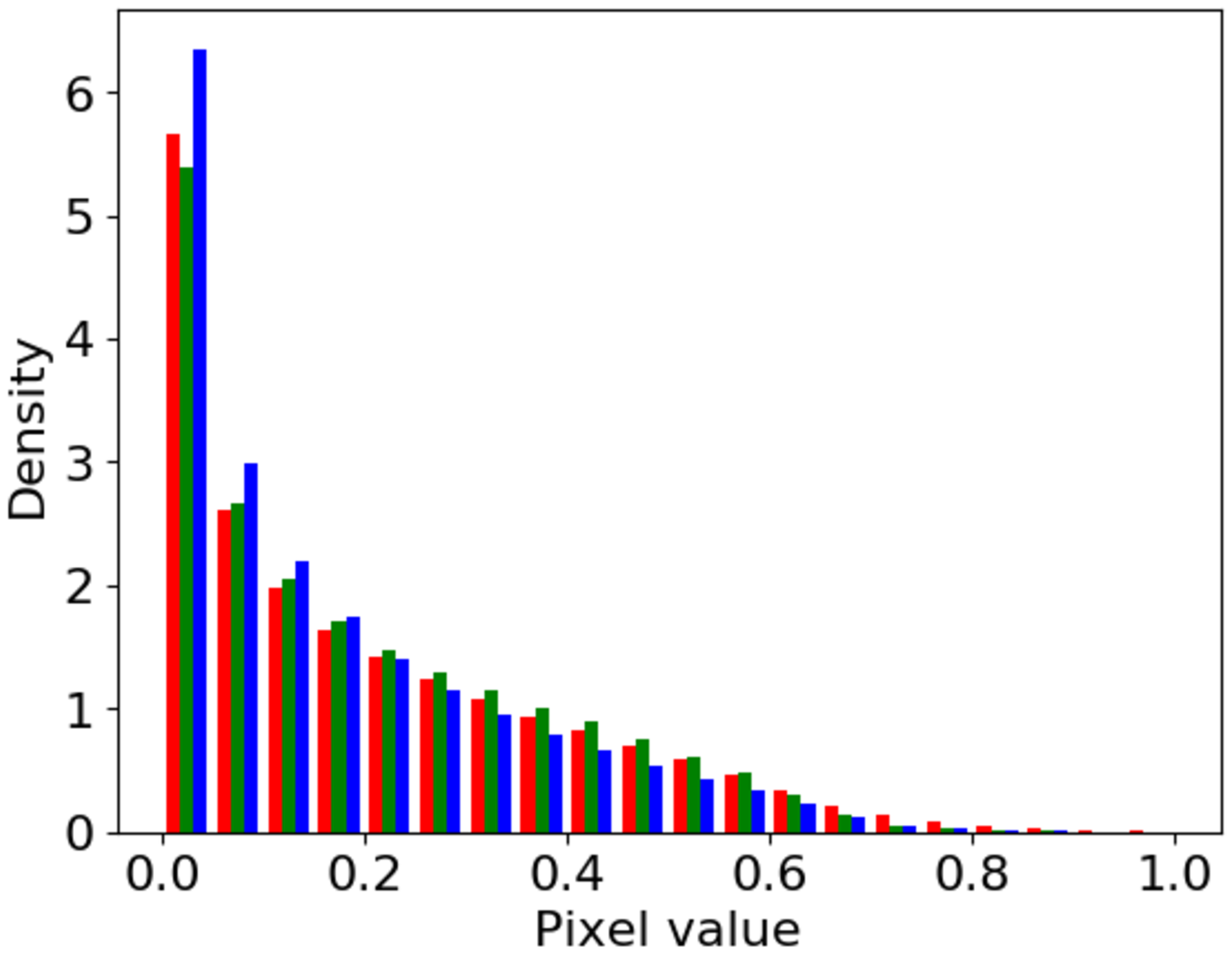}}
\subfloat[Original RGB]{\centering\label{fig:rgb-hist}\includegraphics[width=0.32\textwidth]{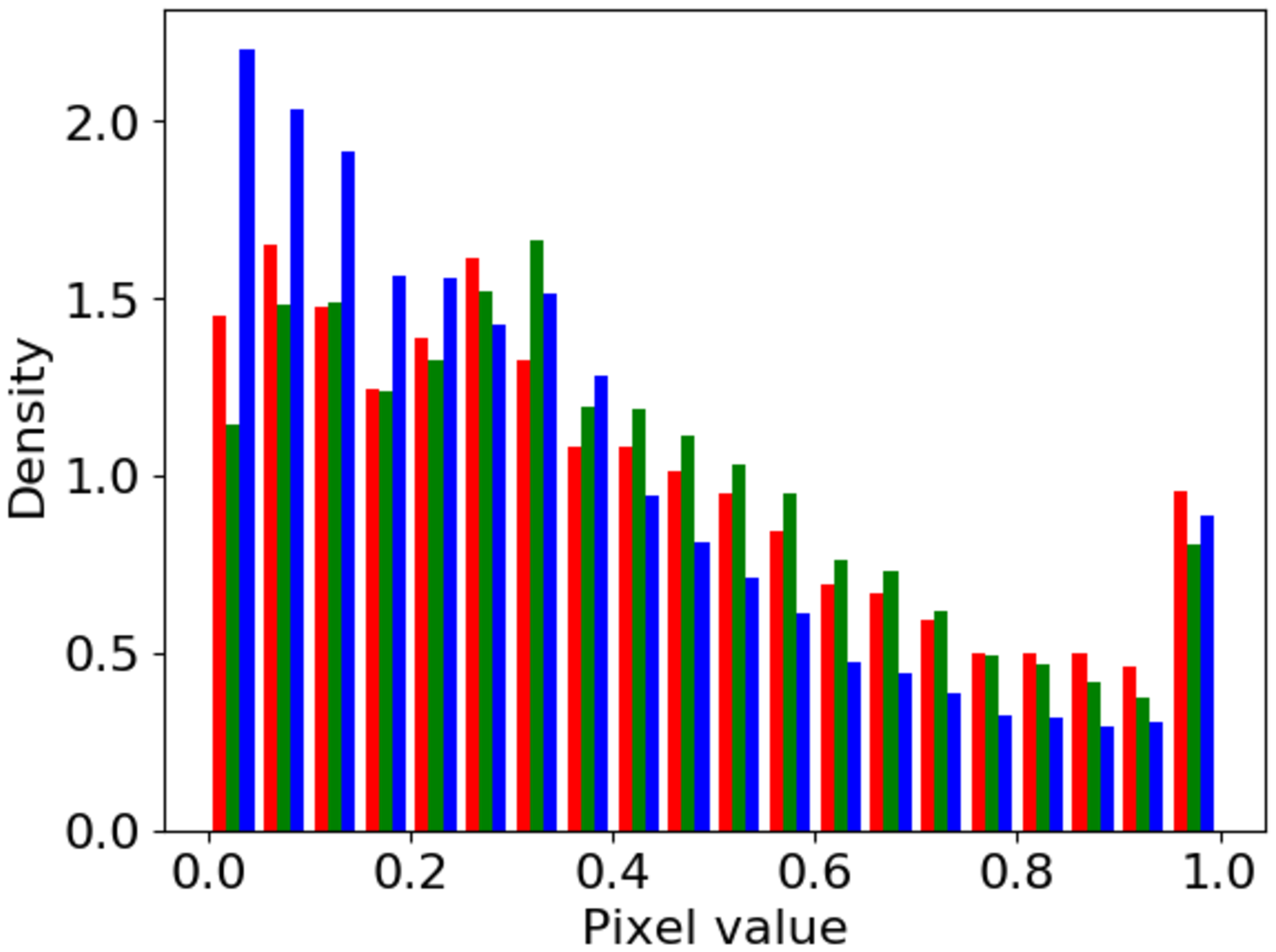}}
\caption{Pixel value distributions (by color channel) at three points along the ISP pipeline. Sampled data is comprised of 25 random images from each class in the ImageNet training set.}
\label{fig:hist}
\end{figure*}

\subsection{Ablation of ISP Algorithms}

Having demonstrated a benefit of ISP processing, we aim to understand which components of the ISP contribute most to improved classification accuracy. To this end, we train MobileNets on images generated with different stages of the ISP selectively enabled. Because of the prevalence of research into the impact of noise on neural networks, the following ablation study is separated into (1) a study of only denoise and (2) a study of remaining ISP components. 

The denoising stage of an ISP is designed to reduce the measurement noise that exists in a raw image due to physical characteristics of the sensor.
Diamond et al.~\cite{dirty-pixels} demonstrate poor compatibility between traditional denoising algorithms and CNNs inference. Our results (Figure \ref{fig:denoise}) agree with this conclusion; we observe an average accuracy improvement provided by denoising of 0.27\% and 0.09\% (within the bounds of model variance) for raw images and RGB images, respectively. In several cases, models trained on denoised images even performed \textit{worse} than their noisy counterparts. It is possible that the denoiser, while removing noise, also tends to blur fine detail which may have otherwise been useful to the CNN. These results indicate negligible impact of traditional denoising on classification accuracy and questionable value for including a denoiser in a CV system.

To evaluate the remaining ISP stages, we compare the performance of ISP configurations with incrementally more stages added until arriving at a full ISP pipeline\footnote{Considering only valid configurations, e.g., tone mapping should only be performed after black level subtraction so that the black level offset is not scaled with light intensity.}. We chose two configurations to evaluate and compare classification performance: (1) black level subtraction (BL) + tone mapping (TM) and (2) black level subtraction (BL) + white balance (WB) + tone mapping (TM). Figure \ref{fig:partial} shows the result of models trained on data processed with these ISP configurations, along with baselines (models trained on raw images and full RGB images). For each stage added, we see some improvement in accuracy across all models. This indicates that each of these ISP stages contributes to improved CNN performance. The most significant improvement in accuracy (5.8\% improvement) is between models trained on raw images and models trained on images processed using black level subtraction and tone mapping. This improvement in accuracy should be attributed to tone mapping because black level subtraction merely shifts the input pixel values by a constant, whereas tone mapping has a significant impact on the statistics of input images. We explore this phenomenon in greater detail in Section \ref{sec:tone-mapping}.

\subsection{Capture Model Validation} \label{sec:validation}
Lubana et al. questioned the use of simulated raw data for training CNN models~\cite{lubana}. Therefore, we validate that our results generalize to real data by using real raw image captures to test CNN models.
Since the capture model does not perfectly reverse ISP processing, it cannot be validated by simply comparing a raw image to the result of running that image through our ISP model and then the capture model. 
Instead, we want to show that the capture model does not introduce any artifacts that negatively impact CNN prediction accuracy on real image captures.
To that end, we created a new dataset of real raw images with which we test our CNN models. This dataset contains 4000 raw images of 50 objects, taken with a variety of lighting conditions and exposure settings. The images are captured using an IMX290 sensor~\cite{imx290}.
We process these images using our ISP model to generate images with which to test our CNN models previously trained on simulated raw images. Note that real raw images processed using certain ISP settings are only tested on models trained with simulated raw images processed using those same ISP settings.

Testing our trained models on this lab captured validation set provides results that broadly follow the trends of Figures \ref{fig:raw-vs-rgb} and \ref{fig:ablation}.
The overall test accuracy on our lab-captured validation set is lower than on the processed ImageNet validation set due to more difficult lighting conditions and levels of noise. However, the differences in accuracy between different ISP configurations is similar to those in our results from simulated raw images. The average accuracy difference between these no ISP and full ISP processing on these real raw images is 5.0\% (compared to 7.0\% on images simulated by the capture model). The average accuracy improvement provided by tone mapping on this dataset is 3.1\% (compared to 5.8\%).
The agreement between results from real and simulated raw images helps validate our approach.
\section{Understanding the Role of Tone Mapping}
\label{sec:tone-mapping}

Having identified tone mapping as a significant component for achieving high CNN accuracy, we next investigate the effect the tone mapper has on the statistics of input images, and how that influences classification performance.

\subsection{Histogram Analysis}
The tone mapper we implement in our software ISP model performs localized histogram equalization\footnote{Our tone mapper compensates for gamma correction later in the ISP pipeline, resulting in a higher concentration of values toward zero.} to pixel values. Therefore, tone mapping tends to result in images that better occupy the full dynamic range, as seen in Figure \ref{fig:hist}, which illustrates the histograms of ImageNet data at three points in the image processing pipeline.


The distribution of simulated raw pixel values (Figure \ref{fig:raw-hist}) differs drastically from that of RGB images (Figure \ref{fig:rgb-hist}). The distribution of raw images concentrates heavily near the black level value of the sensor, while larger values are very sparse. In contrast, the RGB distribution is relatively uniform along its full range [0,1]. The non-uniformity of the raw distribution is exacerbated due to the HDR conditions introduced by the capture model, which increases the overall contrast in the image. 
More formally, we consider the \textit{skew} and \textit{kurtosis}, which describe the asymmetry and peakedness of a distribution, respectively.
The raw distribution has much higher skew and kurtosis than RGB images (Figure \ref{fig:hist}).
Tone mapping does much of the work in transforming the distribution of raw pixel values to that of RGB, as can be seen in Figure \ref{fig:iridix-hist}. The skew and kurtosis of this distribution are quite close to that of RGB images when compared against raw images.

\subsection{Testing our Distribution Hypothesis}
We hypothesize that the difference in image distributions we observe between raw and RGB images (attributed to the tone mapper) results in improved test accuracy for models trained with RGB data.
Lubana et al.~\cite{lubana} showed that a difference in image pixel distributions between training and testing may cause a drop in test accuracy.
However, in this work, we consider the impact the pixel distributions constant between training and test time, but different across experiments.
To explore this, 
we trained CNN models using image data with a range of skew and kurtosis values. We produce such distributions by applying a pixel-wise transform (Equation \ref{eq:f}) to input images, where each pixel of an image, $x$, is normalized to the range $[0,1]$, and $n$ controls the degree of skew/kurtosis.
\begin{equation}
    f(x) = x^n \label{eq:f}
\end{equation}
\begin{equation}
    f^{-1}(x) = \sqrt[\leftroot{-3}\uproot{3}n]{x} \label{eq:finv}
\end{equation}

This transformation shifts data toward low values and creates a distribution resembling that of raw images. Note that no information is being destroyed via this transform, and the original image can be perfectly\footnote{Ignoring minimal numerical loss due to floating-point arithmetic.} reconstructed by the inverse transform (Equation \ref{eq:finv}). Therefore, only the change in input data distributions can cause model accuracy degradation.

We trained ResNet-18~\cite{resnet} on CIFAR-10~\cite{cifar10} using this transformation on both the training and test images. Training on CIFAR-10 enables us to reasonably train many models to build confidence in our results. We sweep $n$ across the range $[1,10]$ with increment $0.5$ and correspondingly generate data with which to train ResNet-18. 

\begin{figure}[t]
\centering
\includegraphics[width=0.48\textwidth]{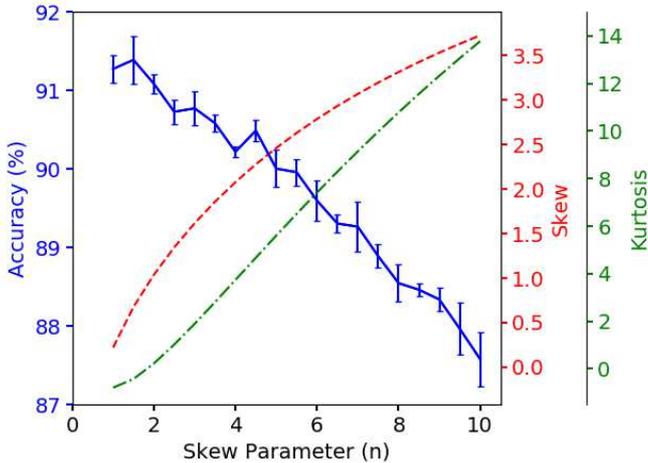}
\caption{Accuracy of ResNet-18 model trained with CIFAR-10 dataset processed with $f(x)=x^n$ transformation. Mean and standard deviation of 5 runs for each value of $n$ are depicted in blue. Skew and kurtosis of the pixel distributions are shown in red and green.}
\label{fig:skew-acc}
\end{figure}

Figure \ref{fig:skew-acc} shows that the distribution of the input image has a considerable impact on accuracy.
Image distributions with high skew result in trained model accuracy with considerable accuracy reduction relative to baseline ($n=1$).
At $n=10$, there is an average accuracy degradation of 3.70\%. 
Even with such a drastic change to the input distribution, the skew and kurtosis of the input data in this experiment (skew = 3.67, kurtosis = 13.28) are much lower than that of our simulated raw ImageNet data (skew = 7.85, kurtosis = 95.80). Therefore, it is not surprising that using tone mapping to normalize the input data distribution results in models with an average of 5.77\% higher accuracy than those trained on raw data.
\section{System Level Impact of ISP}
\label{sec:system}

In this section, we examine the practical system level impact of the ISP on: 1) the accuracy vs hardware cost trade-offs, and 2) generalization to different imaging sensors.

\subsection{Hardware Efficiency Trade-offs}
We compare hardware cost in terms of two critical metrics: the arithmetic operations per inference, and the total memory accessed per inference. The number of operations per input image is calculated for both our ISP model and MobileNets\footnote{Operations are counted as multiply, add, and simple transcendental functions. Multiply-accumulate (MAC) is counted as two operations.}. For memory, we assume the worst case scenario for the ISP -- the ISP must read the input image from DRAM rather than streaming directly from the image sensor.

\begin{figure}[t]
\centering
\vspace{-3mm}
\includegraphics[width=0.45\textwidth]{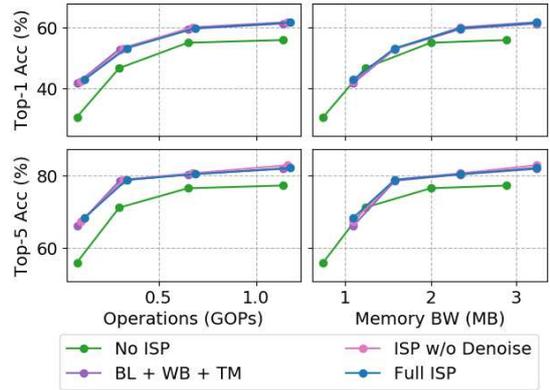}
\caption{Compute complexity (GOPs/inference) and memory cost (MB/inference) vs prediction accuracy trade-offs for MobileNets with inputs processed by different ISP configurations.}
\label{fig:complexity}
\end{figure}

Figure \ref{fig:complexity} illustrates the trade-offs between hardware cost and accuracy for different ISP configurations and MobileNet widths. The relative improvement in accuracy provided by a minimally-viable ISP (BL+WB+TM) far outweighs the additional hardware cost. Additional ISP complexity provides both marginal increases in accuracy and marginal computational costs. However the improvements in accuracy prove worthwhile because the cost of even a full ISP is minuscule in comparison to any relevant CNN architecture. Therefore, at iso-accuracy, it is clear that a full ISP pipeline provides enough benefit to warrant its cost.

\subsection{Model Compatibility and Generalization}
Beyond improving computational efficiency, the ISP significantly eases system deployment. One problem we quickly encountered with training CNNs on raw images is that many networks, especially object detection networks, require images to be resized to an expected shape. However, resizing raw images is not trivial because color information is embedded in a single plane. therefore, this prohibits raw images from being used with many relevant networks and CV pipelines.

\begin{table}[t]
\centering
\resizebox{\columnwidth}{!}{
\begin{tabular}{c c c}
\hline
 Image Processing & \multicolumn{2}{c}{Classification Accuracy (\%)} \\
 & Sensor A \cite{ar231} & Sensor B \cite{imx290} \\

\hline\hline
No ISP (raw) & 53.9 & 31.7 \\
Full ISP (RGB) & 56.5 & 52.2 \\
\hline
\end{tabular}
}
\caption{Generalization of MobileNet-1.0 across sensors. 
}
\label{table:camera-abstraction}
\end{table}

Another benefit of using an ISP is that RGB images are a standard representation that is somewhat invariant to the sensor used to capture the image. Raw images may have drastically different statistics depending on what type of sensor is used, and even due to device variation from the manufacturing process. To demonstrate the impact of an ISP on the generalization of a CNN to multiple sensors, we present the following experiment: we train MobileNet-1.0 with images captured by Sensor A (AR231 \cite{ar231}), and test on images captured by Sensor B (IMX290 \cite{imx290}). Using ImageNet processed by a capture model and an ISP tuned for each sensor, we found that a system with an ISP loses only 4.3\% test accuracy after switching sensors (Table \ref{table:camera-abstraction}). In contrast, a system with no ISP incurs a huge 21.8\% accuracy loss. Such accuracy losses could be mitigated by on-device learning, however that is arguably currently impractical due to the high computation costs of training CNNs. Therefore, an ISP provides a significant benefit to the generalization of a CNN to deployment in different hardware generations.
\section{Conclusion}
\label{sec:conclusion}

In this paper, we empirically studied the impact of image signal processing (ISP) on CNN prediction accuracy.
This was performed using a software model of an ISP and a model of an imaging sensor to enable the study of relevant application domains. 
We validated the approach by comparing training results from simulated raw images against raw images captured in-lab. 
We found that processing images with an ISP improves accuracy by an average of 7.0\% for a broad set of MobileNet CNNs. Our results indicate that the ISP has a more significant impact on smaller CNN models, and our results on the larger ResNet-50 and ResNet-101 models are consistent with this trend.
Each component of the ISP pipeline provides accuracy gains across all models, except for denoise, which is found to have questionable benefit to CNN performance.
Tone mapping, which equalizes pixel value distributions in our implementation, provides a significant 5.8\% accuracy improvement. 
We show empirical evidence that uneven pixel distributions arising from raw images are the cause of degraded classification performance. 
The ISP also allows CNNs to generalize to multiple imaging sensors with significantly reduced (17.5\%) accuracy degradation compared to raw images.
Finally, the ISP benefits system efficiency because the hardware cost is significantly lower than the cost of using a larger CNN to achieve the same accuracy.

{\small
\bibliographystyle{ieee_fullname}
\bibliography{root}
}

\clearpage
\appendix

\begin{figure*}[ht]
\centering
\subfloat[RGB train accuracy]{\centering\label{fig:rgb-train}\includegraphics[width=0.45\textwidth]{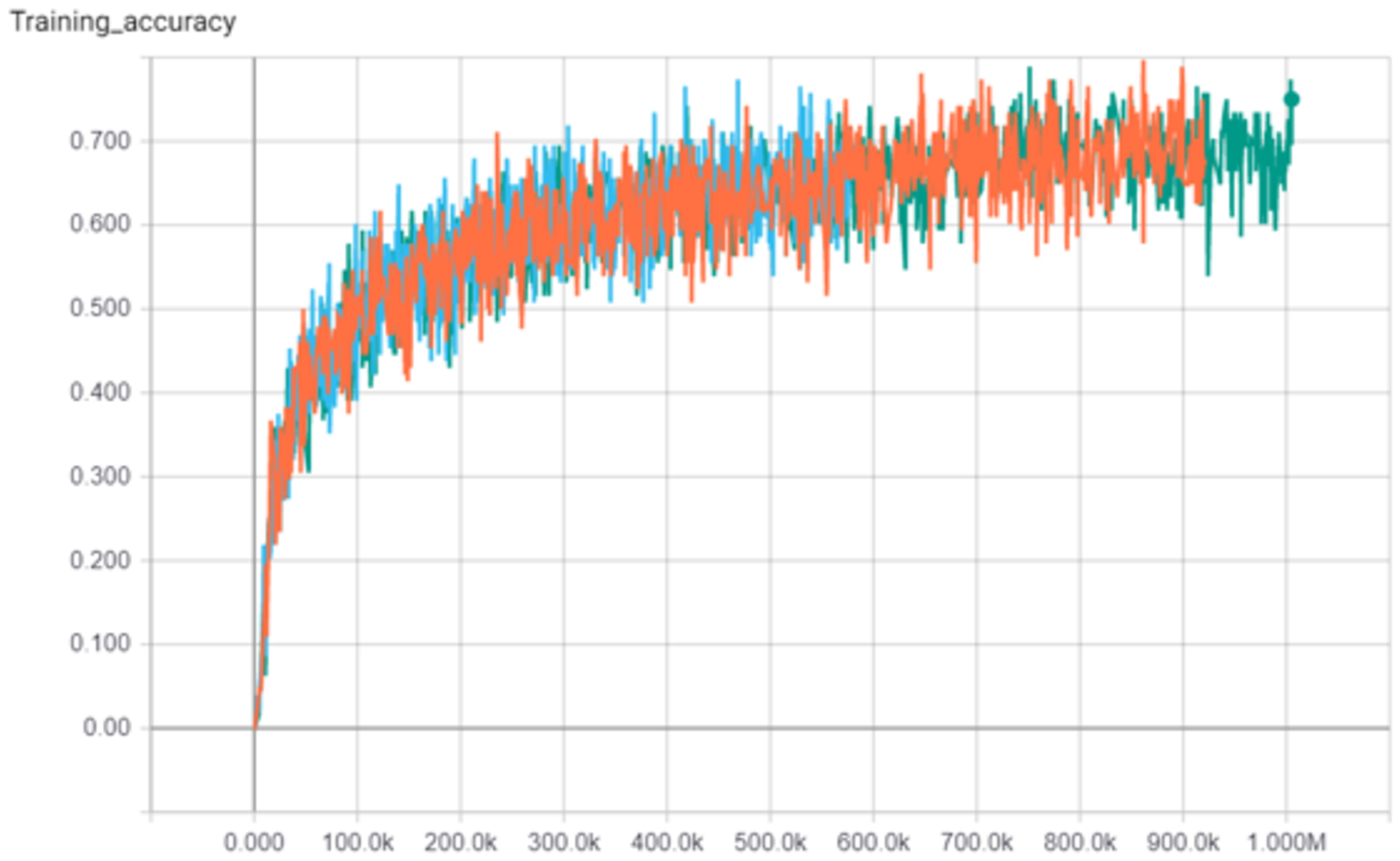}}
\subfloat[RGB validation accuracy]{\centering\label{fig:rgb-val}\includegraphics[width=0.45\textwidth]{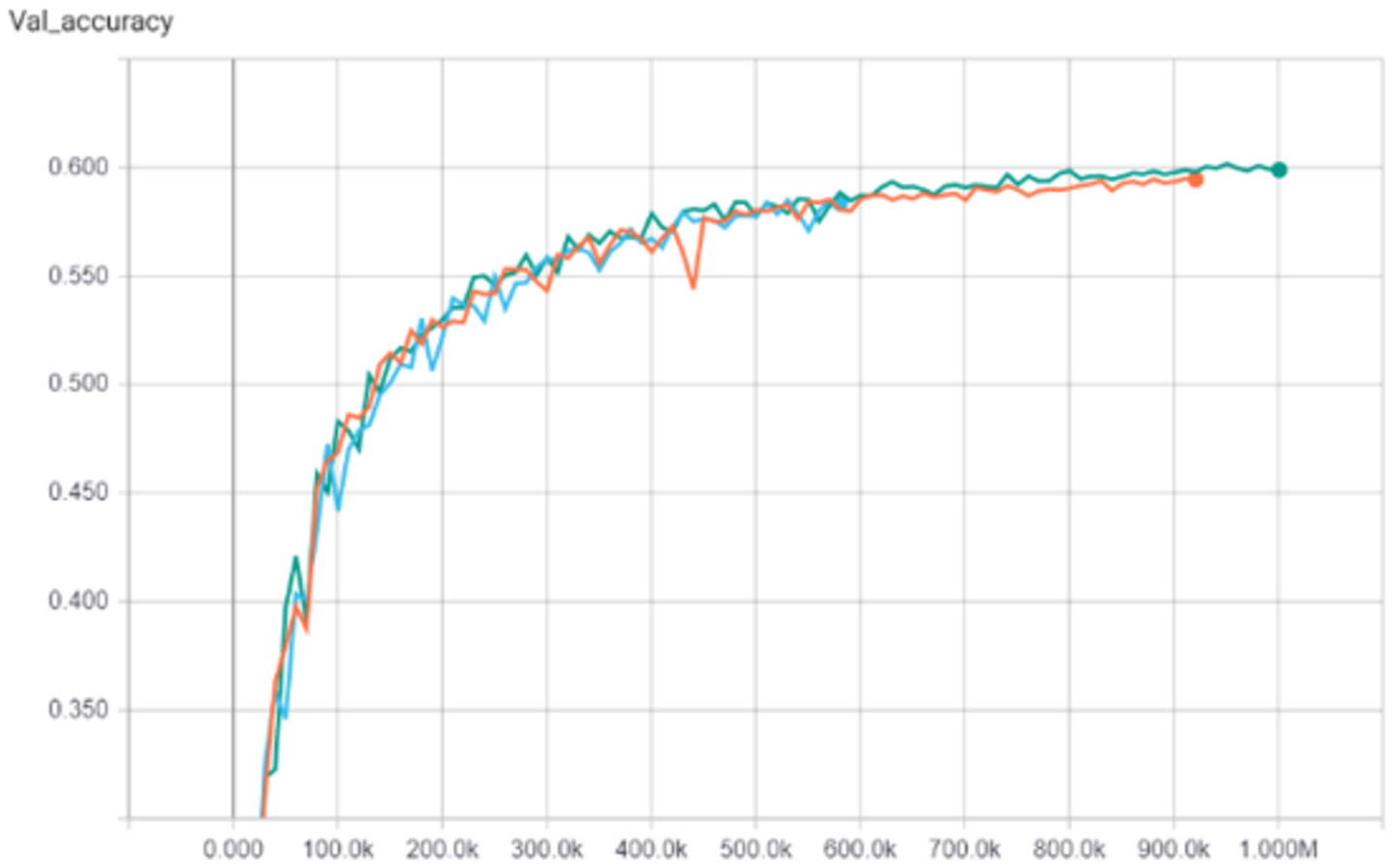}} %

\subfloat[Raw train accuracy with default training recipe]{\centering\label{fig:raw-train}\includegraphics[width=0.45\textwidth]{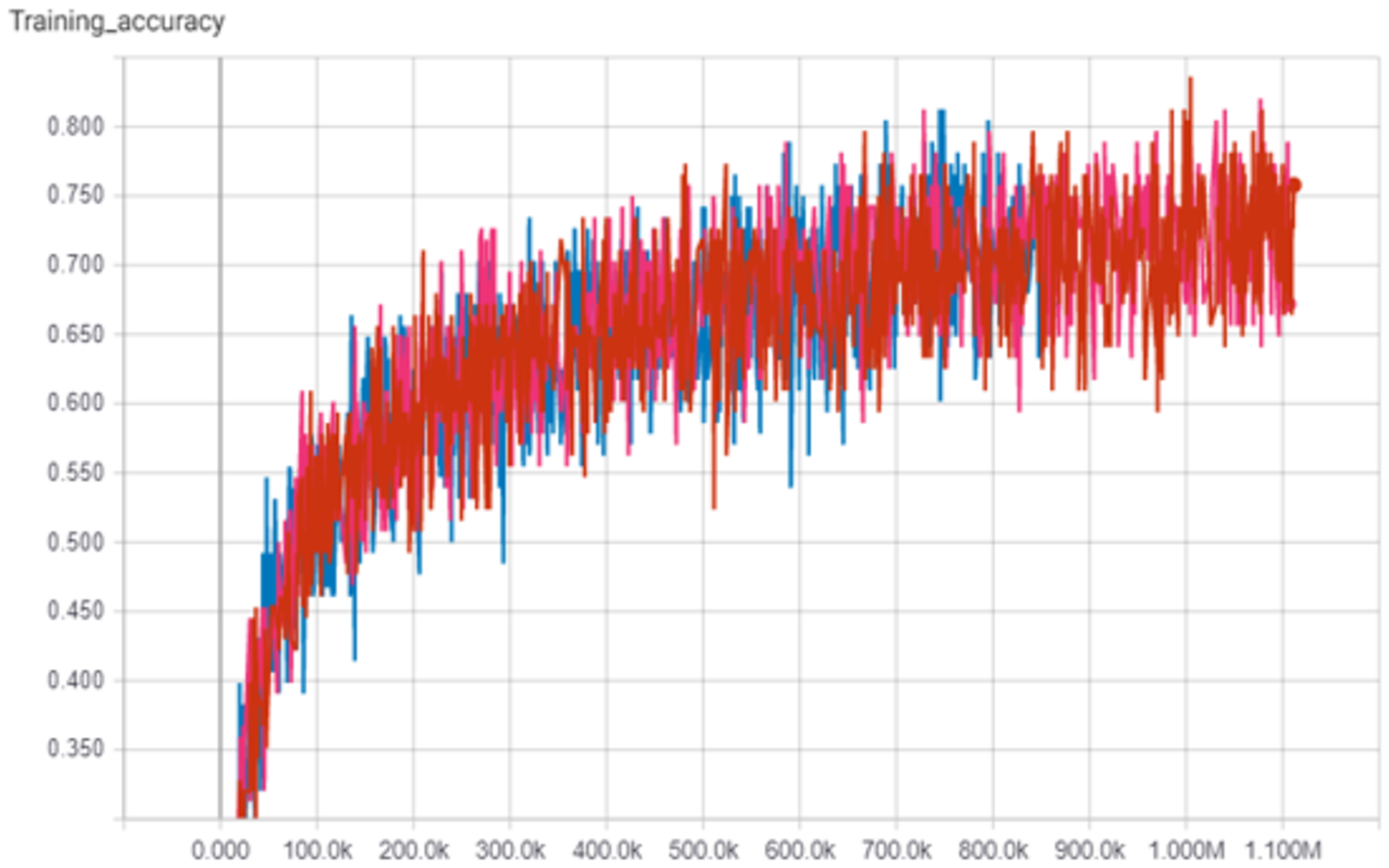}}
\subfloat[Raw validation accuracy with default training recipe]{\centering\label{fig:raw-val}\includegraphics[width=0.45\textwidth]{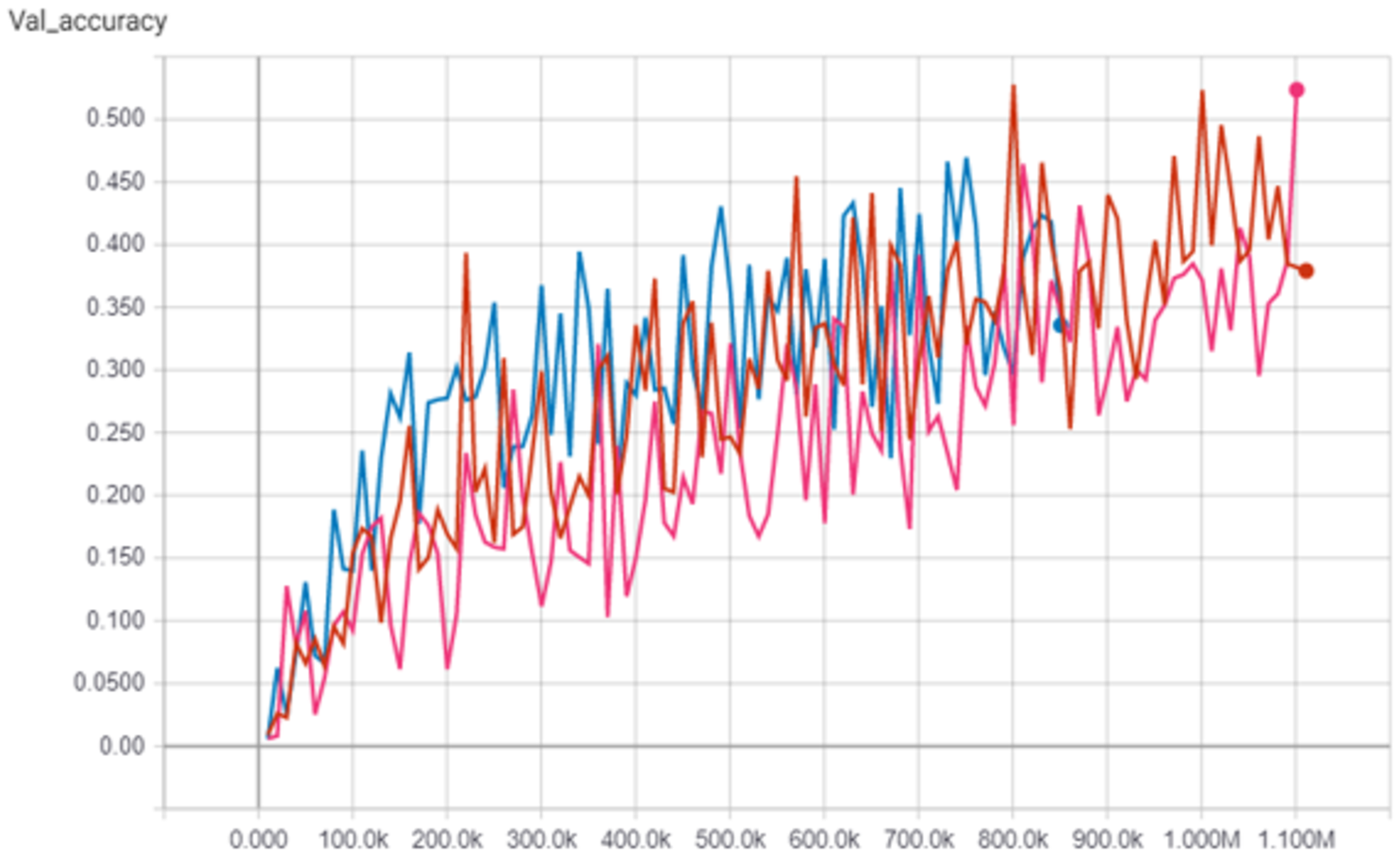}}
\caption{Training and validation accuracy curves for 3 random initializations of MobileNet using default training hyper-parameters.}
\label{fig:curves}
\end{figure*}

\section{Raw Image Representation} \label{sec:representation}
The mapping of mosaic images\footnote{Images not processed with the demosaic ISP stage, i.e. still in a Bayer pattern.} to inputs of traditional CNNs is not obvious because mosaic images images have shape $[H, W, 1]$ whereas the expected input to a CNN has shape $[H, W, 3]$. We considered three different representations of mosaic images to be used for training CNNs:
\begin{enumerate}
    \item Keep pixels on a single plane (shape = $[H, W, 1]$)
    \item Stretch pixels to 3 channels based on color, and insert zeros for missing color information (shape = $[H, W, 3]$)
    \item Stretch pixels to 4 channels based on Bayer pattern (R, Gr, Gb, B), and do not insert zeros (shape = $[H/2, W/2, 4]$)
\end{enumerate}

Each of these representations can be used as inputs to MobileNets and ResNets with little or no modification to the first convolutional layer. Representation 1 requires the first convolution layer to have 1 input feature map. Representation 3 requires the first convolution layer to have 4 input feature maps and a stride of 1 (so that the output activations have the same shape as the baseline model). Representations 1 and 2 have information of the Bayer pattern embedded in 2x2 spacial grids, but luckily the first layer in each network has a stride of 2 so each weight corresponds to an individual color in the Bayer pattern.

We trained 3 copies of MobileNet on raw data using each of the representations, and found negligible difference in performance amongst the models. We chose to publish results using Representation 2 because it requires no modifications to the CNNs, which we felt was the most fair approach when comparing to models trained on demosaiced images.

\section{Training on Raw Images} \label{sec:raw-training}
Figure \ref{fig:curves} depicts the training and validation top-1 accuracy for several MobileNets during training. These models were all trained with the default hyperparameters from the original publication of MobileNets \cite{mobilenetv1}. We noticed that accuracy curves from models trained on raw images (Figures \ref{fig:raw-train} and \ref{fig:raw-val}) had different shape than those from models trained on RGB images (Figures \ref{fig:rgb-train} and \ref{fig:rgb-val}). The accuracy curves associated with raw images have much more variation between each iteration, indicating instability of the optimizer.
We recognize that the training recipe we used was tuned assuming RGB input images, and the performance clearly does not translate well to raw images. We intuited that the cause for the poor accuracy on raw images was batch normalization, which behaved poorly when used with input images with pixel distributions seen in raw images. Upon investigation of batch normalization hyper-parameters, we determined that changing batch normalization decay from 0.9997 to 0.99 removed instability from the raw training curves. This change effectively enables batch normalization statistics to update more quickly, which we believe is important for raw data due to its highly non-linear pixel distribution. The described change to the raw training recipe successfully removed the variation in accuracy seen in Figures \ref{fig:raw-train} and \ref{fig:raw-val}. We explored other changes to the raw training recipe, but we found that none of our tests yielded better results.

\begin{table*}[t]
\centering
\begin{tabular}{c c c c c}
\hline
Image processing & MN-0.25 acc. (\%) & MN-0.50 acc. (\%) & MN-0.75 acc. (\%) & MN-1.00 acc. (\%)  \\
\hline\hline
None & 1.70 & 6.70 & 9.50 & 7.65 \\
Denoise & 1.54 & 7.20 & 8.00 & 8.43 \\
BL + TM & 2.65 & 8.67 & 7.54 & 11.33 \\
BL + WB + TM & 1.95 & 8.77 & 8.02 & 10.96 \\
ISP w/o denoise & 2.45 & 8.50 & 8.35 & 14.00\\
Full ISP & 3.15 & 9.90 & 8.65 & 14.95 \\
\hline
\end{tabular}
\caption{Top-1 test accuracy on real data for MobileNets trained on simulated data (MN = MobileNet).
}
\label{table:validation-top-1}

\end{table*}
\begin{table*}[t]
\centering
\begin{tabular}{c c c c c}
\hline
Image processing & MN-0.25 acc. (\%) & MN-0.50 acc. (\%) & MN-0.75 acc. (\%) & MN-1.00 acc. (\%)  \\
\hline\hline
None & 5.00 & 19.15 & 21.10 & 21.35 \\
Denoise & 3.35 & 17.85 & 20.95 & 22.80 \\
BL + TM & 9.23 & 20.10 & 21.90 & 26.05 \\
BL + WB + TM & 10.54 & 24.01 & 22.50 & 27.20 \\
ISP w/o denoise & 10.40 & 23.60 & 20.65 & 26.15 \\
Full ISP & 13.10 & 27.60 & 22.35 & 29.20 \\
\hline
\end{tabular}
\caption{Top-5 test accuracy on real data for MobileNets trained on simulated data (MN = MobileNet).
}
\label{table:validation-top-5}
\end{table*}

\begin{figure*}[h]
\centering
\subfloat[ImageNet samples]{\centering\label{fig:rgb-train-images}\includegraphics[width=0.45\textwidth]{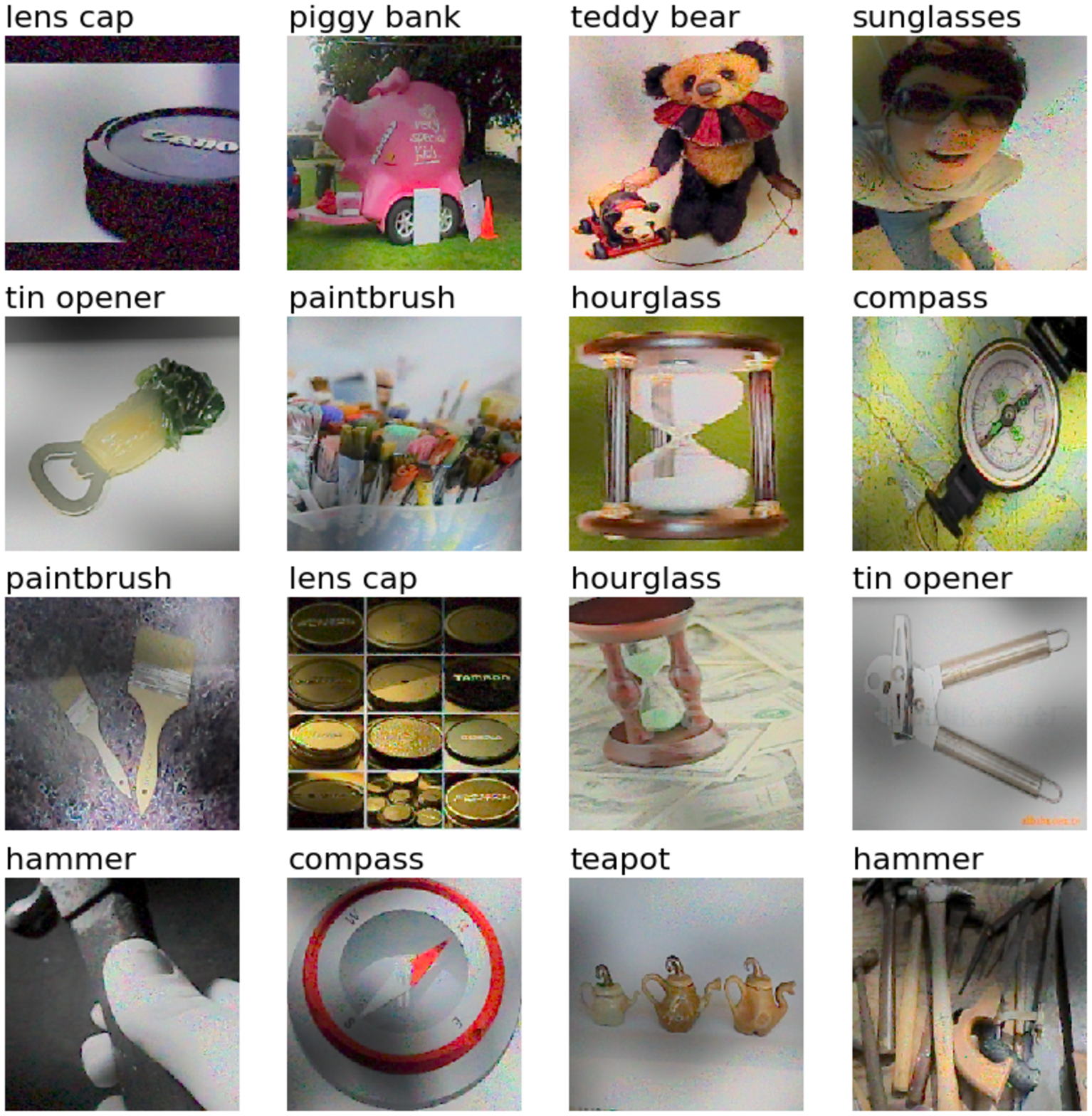}}
\subfloat[Lab-captured samples]{\centering\label{fig:rgb-val-images}\includegraphics[width=0.45\textwidth]{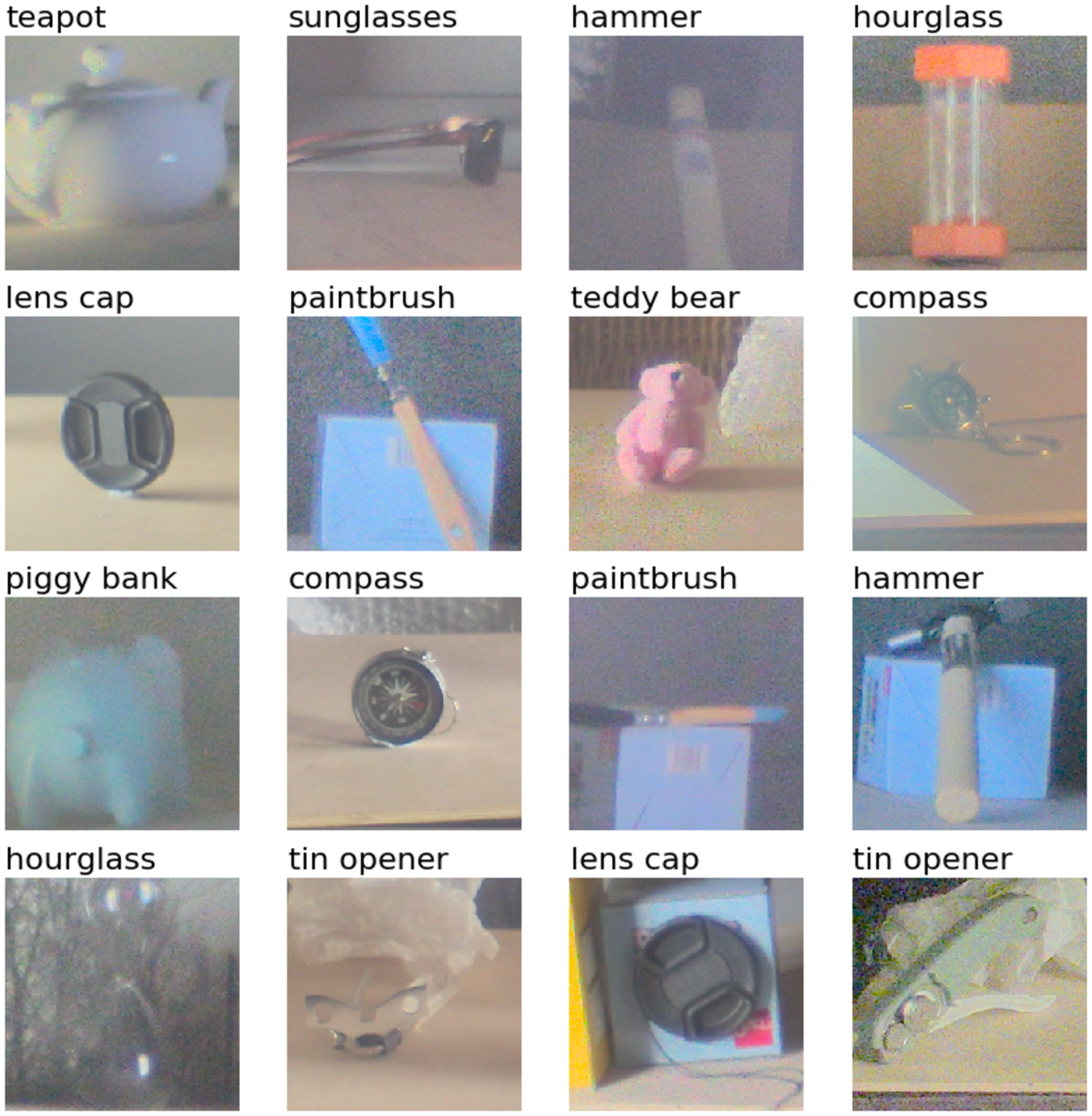}}
\caption{Selection of images from ImageNet and from our lab-captured dataset (sampled randomly). The images shown from ImageNet have been processed using our capture model and ISP model, whereas the images shown from our lab-captured dataset have only been processed using our ISP model.}
\label{fig:samples}
\end{figure*}

\section{Lab-Captured Validation Set Results} \label{sec:validation-results}
Section \ref{sec:validation} discusses validation tests designed to provide confidence in our experimental methodology (specifically, to ensure that the capture model does not introduce any strange behavior when used for CNN training). Tables \ref{table:validation-top-1} and \ref{table:validation-top-5} provide the accuracy of testing the models trained on data simulated using the capture model. We see similar trends in these test results as we do when testing on processed versions of the ImageNet validation set (Figures \ref{fig:raw-vs-rgb} and \ref{fig:ablation}).

The overall test accuracy on our lab-captured validation set is lower than on the processed ImageNet validation set. We believe that the difference in accuracy is due to a higher difficulty to classify the images in our lab-captured dataset. The levels of noise in our dataset are much higher, and lighting conditions are on average worse than what is found in ImageNet. These differences can be seen in Figure \ref{fig:samples}, which displays randomly sampled images from the ImageNet validation set and our lab-captured dataset.

\end{document}